\documentclass[pre,twocolumn,showpacs]{revtex4}
\usepackage[xdvi]{graphicx}%
\usepackage{dcolumn}
\usepackage{amsmath}
\makeatletter

\def\btt#1{\texttt{\@backslashchar#1}}%
\DeclareRobustCommand\bblash{\btt{\@backslashchar}}%
\makeatother

\newcommand{\bra}{\left\langle}
\newcommand{\ket}{\right\rangle}

\newcommand{\bv}[1]{{\boldsymbol #1}}

\newcommand{\Rc}{\nu_{\rm c}}

\begin{document}

\title{Criticality and Scaling Relations in a Sheared Granular Material}

\author{Takahiro Hatano}
\email[Electronic address: ]{hatano@eri.u-tokyo.ac.jp}
\affiliation{Earthquake Research Institute, University of Tokyo, 113-0032, Japan}

\author{Michio Otsuki}
\email[Electronic address: ]{otsuki@jiro.c.u-tokyo.ac.jp}
\affiliation{Department of Pure and Applied Sciences, University of Tokyo, 
Komaba, Tokyo 153-8902, Japan}

\author{Shin-ichi Sasa}
\email[Electronic address: ]{sasa@jiro.c.u-tokyo.ac.jp}
\affiliation{Department of Pure and Applied Sciences, University of Tokyo, 
Komaba, Tokyo 153-8902, Japan}

\date{\today}

\begin{abstract}
We investigate a rheological property of a dense granular material under shear. 
By a numerical experiment of the system with constant volume, we find 
a critical volume fraction at which the shear stress and the pressure 
behave as power-law functions of the shear strain rate. We also present 
a simple scaling argument that determines the power-law exponents.
Using these results, we  interpret a power-law behavior observed in 
the system under constant pressure. 
\end{abstract}


\maketitle


\section{introduction}

Soft glassy systems such as foams, colloidal suspensions,  
emulsions, polymers, glasses \cite{glass1,glass2,glass3},
and granular materials 
\cite{Nagel,review,granular} have a strongly non-linear response to 
an external perturbation. In such systems, the relation between the 
stress $\sigma$ and the strain rate $\gamma$ characterizes the system
behavior.  Although  it is known that the relations are diverse
 and specific 
to individual systems, a universal law for a certain class of systems
may exist.

\begin{figure}[htbp]
\begin{center}
\includegraphics[height=15em]{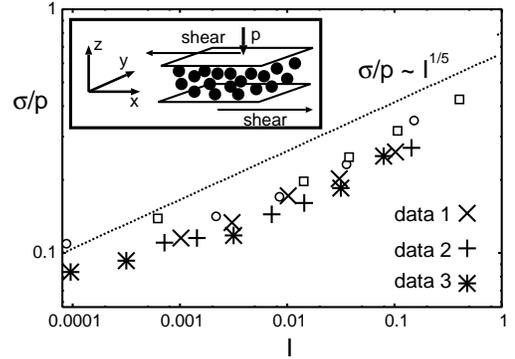}
\caption{$\sigma/p$ as a function of $I$. This result was obtained 
for a model similar to that explained in the text. The main 
differences are as follows: (i)  the top boundary in the $z$ direction
is modified so as to maintain a constant pressure and (ii) the shear is applied 
directly from the moving layer at the top and the bottom. (See the inset.) 
The parameter values are as follows: $L_x/a=L_y/a=25$, $N=10800$, and 
$\eta/\sqrt{km}=1.0$. $pa/k=1.92 \times 10^{-3}$ (data 1), 
$3.75\times 10^{-5}$ (data 2), and  $pa/k=4.35 \times 10^{-3}$ 
with $L_x/a=L_y/a=15$ (data 3). Furthermore, the square and
circle symbols represent the constant pressure data  obtained 
from Figs. \ref{fig:sg}  and  \ref{fig:pg}, where 
$pa/k=\Pi=1.25\times 10^{-3}$ (square symbol) and $5.4\times 10^{-5}$ 
(circle symbol).
}
\label{fig:hatano}
\end{center}
\end{figure}

In particular, in sheared granular materials under constant pressure  $p$,
one of the authors (Hatano) has found a relation \cite{hatano}
\begin{equation}
\frac{\sigma}{p} \simeq I^\phi
\label{hatano:power}
\end{equation}
with 
\begin{equation}
I= \sqrt{\frac{m}{pa}}  \gamma
\label{I:def}
\end{equation}
by a numerical experiment using the discrete element method. 
Here, $a$ is the maximum diameter of the particles (their diameters are 
uniformly distributed in the range $[0.7a, a]$) and $m$ is the mass 
of the particles \cite{fn:hatano}. As demonstrated in Fig. \ref{fig:hatano},
the exponent $\phi$ is not inconsistent with  $1/5$ in the range 
$ 10^{-3} \le I \le 10^{-1}$. Surprisingly, the power-law behavior 
given in Eq. (\ref{hatano:power}) is observed in the cases that 
$p/Y \simeq 10^{-3}$ and $10^{-5}$, where $Y$ represents
the Young modulus of the particle. For example, one can experimentally
obtain the power-law behavior under the constant pressure 
$p=10^0-10^{-2}$MPa by using  polystyrene with $Y= 3$GPa. Since 
$I=10^{-3}$ corresponds to the shear rate $10^0-10^1$/sec in this 
example, the shear condition leading to Eq. (\ref{hatano:power}) 
is experimentally possible.

Stimulated by this result, in the present paper,
we consider the power-law behavior of stress-strain rate relations
in sheared granular materials by investigating a model granular 
system with the Lees-Edwards boundary conditions. In this idealized 
system, we demonstrate that there is a critical volume fraction 
at which the shear stress and the pressure (normal stress) behave 
as  power-law functions of the shear strain rate
in the limit $\gamma \to 0$. From these power-law behaviors, we derive 
the scaling relation $\sigma/p \simeq I^{1/5}$ in the limit $I \to 0$ 
at the critical volume fraction. Note that this critical condition  
does {\it not} 
correspond to a constant pressure. We then present a simple interpretation
of Eq. (\ref{hatano:power}) for the system under constant pressure. 


\section{Model and numerical results}


Here, we describe our computational model. The system 
consists of $N$  spheres of mass $m$ in a three-dimensional rectangle 
box whose lengths are $L_x$,  $L_y$, and $ L_z$ along the $x$, $y$, 
and $z$ directions, respectively. In order to realize an average  
velocity gradient $\gamma$ in the $z$ direction and average velocity 
in the $x$ direction, we impose 
the Lees-Edwards boundary conditions \cite{Evans}. The particle 
diameters are $0.7a$, $0.8a$, $0.9a$ and $a$ each of which is
assigned to $N/4$ particles. 
When the distance between two particles is less than the sum of their 
radii, $r_1$ and $r_2$, an interaction force acts on each of them. 
This force comprises an  elastic repulsion force 
$k (\delta r-(r_1+r_2))$ and the viscous dissipation force $\eta \delta v$, 
where $\delta r$ 
and $\delta v$ represent the relative distance and  velocity 
difference of the interacting  particles, respectively. For simplicity, 
we do not consider the tangential force between the interacting
particles. We study the specific case where $L_x/a=L_y/a=L_z/a$, 
$N=1728$ and $\eta/\sqrt{km}=1.0$. The control parameters in this system 
are the volume fraction $\nu\equiv \sum_{i=1}^N \pi a_i^3/(6L_x L_y L_z) $
with the $i$th particle diameter $a_i$,  
and the dimensionless shear rate $\Gamma\equiv \gamma \sqrt{m/k}$. 
We then calculate the dimensionless shear stress $\Sigma=\sigma a/k$ and 
the dimensionless pressure (in the $z$ direction) 
$\Pi=p a/k$. See Ref. \cite{Evans} as the calculation method
for $\Sigma$ and $\Pi$. Note that $k/a$ provides an approximate
value of the Young modulus of particles.

\begin{figure}[htbp]
\begin{center}
\includegraphics[height=15em]{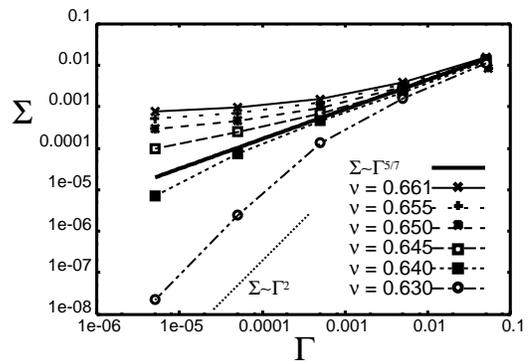}
\caption{$\Sigma$ as a function of $\Gamma$ for several values of $\nu$.
The thick solid line represents $\Sigma \propto \Gamma^{5/7}$ that 
is estimated from our theoretical argument. Note that the Bagnold 
scaling \cite{Bagnold} is observed for the case in which $\nu=0.630$
and $\Gamma \le 10^{-4}$.
}
\label{fig:sg}
\end{center}
\end{figure}
\begin{figure}[htbp]
\begin{center}
\includegraphics[height=15em]{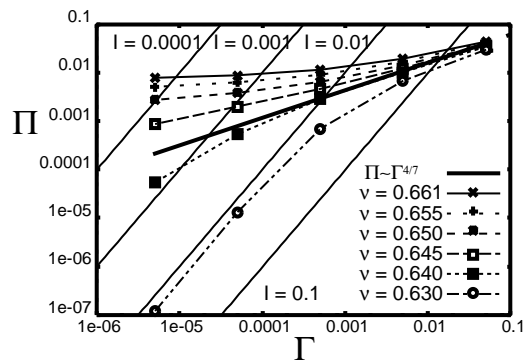}
\caption{
$\Pi$ as a function of $\Gamma$ for several values of $\nu$.
The thick solid line represents $\Pi \propto \Gamma^{4/7}$ 
that is estimated from our theoretical argument. Furthermore, the 
lines 
$I=$ const. are drawn to help us to understand the  functional 
form of $\sigma/p$ over $I$.
}
\label{fig:pg}
\end{center}
\end{figure}

We express the dependence of $\Sigma $ and $\Pi$ on $(\Gamma,\nu)$
as $\Sigma=f_\sigma(\Gamma, \nu)$ and $\Pi=f_p(\Gamma,\nu)$, respectively.
Figures \ref{fig:sg} and \ref{fig:pg} display these functions with respect
to $\Gamma$ for several values of $\nu$ \cite{Campbell}. 
These graphs clearly show that 
there exists a critical volume fraction $\Rc$ at which the power law 
behaviors are observed as follows:
\begin{eqnarray}
f_\sigma(\Gamma,\Rc) &\simeq&  \Gamma^{\alpha} \label{scaling:1}, \\
f_p(\Gamma,\Rc) &\simeq&  \Gamma^{\beta} \label{scaling:2}, 
\end{eqnarray}
in the limit $\Gamma \to 0$ \cite{fn:power2}. 
The values of the exponents will be 
discussed later. 
{ Here, it is worthwhile noting that similar graphs were obtained in 
Ref. \cite{Campbell} 
with the argument on the effect of finite elastic modulus.
Indeed, these graphs in this reference suggest the existence of 
the critical state, although the power-law behavior was not mentioned
explicitly. }
Upon numerical verification, we found that the 
critical volume fraction corresponds to the jamming transition point 
defined as the volume fraction beyond which a finite yield stress 
appears \cite{Nature}. In this paper,
we do not argue the nature of the jamming transition, 
but focus on the power-law behaviors given in Eqs. (\ref{scaling:1}) and 
(\ref{scaling:2}). Note that a similar critical state was obtained
for a sheared glassy system \cite{glass1}. 

\section{Theoretical argument}

\begin{figure}[htbp]
\begin{center}
\includegraphics[height=15em]{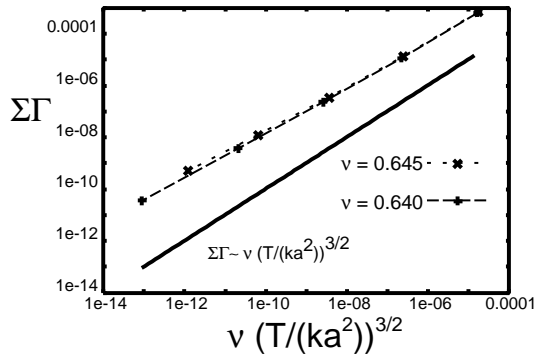}
\caption{
$\nu ({T}/(ka^2))^{3/2}$ versus $\Sigma \Gamma$ for two  cases 
in which $\nu$
is close to  the critical volume fraction $\Rc$. The solid line 
corresponds to Eq. (\ref{eb:dless}). Since we find 
the best fitting to be $\nu ({T}/(ka^2))^{3/2} \sim {(\Sigma \Gamma)}^{0.9}$, 
there is a slight deviation from Eq. (\ref{eb:dless}).
}
\label{fig:Haff}
\end{center}
\end{figure}

The main idea in our theoretical argument is to consider dimensional 
analysis with  kinematic temperature $T$ defined as
\begin{equation}
T= \frac{m}{3N}(\sum_{i=1}^N \bra |\bv{v}_i|^2 \ket- |\bra \bv{v}_i \ket|^2 ),
\end{equation}
where
$\bv{v}_i$ denotes the velocity of the $i$-th particle.
Although $T$ is not a parameter of the system but is determined 
by $\Gamma$ and $\nu$, it is considered that physical processes 
in granular systems are described in terms of the kinematic 
temperature \cite{Mitarai}. 
In particular, { the time scale of energy dissipation 
is assumed to be determined as $(\sqrt{T/m}/a)^{-1}$. }
One can verify the
validity of this assumption by investigating the energy balance equation 
in the steady state \cite{Mitarai}:
\begin{equation}
\frac{N}{L_xL_yL_Z}\frac{T^{3/2}}{a \sqrt{m}}  \simeq \sigma \gamma,
\label{eb}
\end{equation}
which is rewritten as
\begin{equation}
\nu \left(\frac{T}{ka^2}\right)^{3/2} \simeq \Sigma \Gamma.
\label{eb:dless} 
\end{equation}
Figure \ref{fig:Haff} indicates that Eq. (\ref{eb:dless}) 
is plausible as the first theoretical attempt,
although a slight deviation is observed. 
Based on  this result, 
hereafter, we assume that { the time scale of 
the energy dissipation is given by $(\sqrt{T/m}/a)^{-1}$.}

Now, we consider a set of  dimensionless constants using the energy
dissipation ratio. Let us recall that we have three independent 
dimensionless parameters $(\nu,\Gamma, \eta/\sqrt{km})$. 
We want to use dimensionless parameters each of which  represents
the ratio of { a time scale  of a physical process to that of energy 
dissipation}. It is then reasonable that the following { two time scales}
are important: {the inverse of the shear rate} $\gamma^{-1}$ and { the 
relaxation time} $\eta/k$ of the elastic displacement during the 
interaction between two particles. Thus, instead of $\Gamma$ and 
$\eta/\sqrt{km}$, we introduce
\begin{eqnarray}
\xi_1 &=&  a \sqrt{\frac{m}{T}} \gamma,  \label{xi1} \\
\xi_2 &=&  \frac{1}{ka}\sqrt{\frac{T}{m}}\eta. \label{xi2} 
\end{eqnarray}

Then,  we wish to determine the functional forms of 
$\Sigma(\xi_1,\xi_2) $ and $\Pi(\xi_1,\xi_2) $ in the limit $\Gamma \to 0$
at the critical volume fraction. In order to restrict the possible 
forms of the 
functions, we further assume that $\sigma$ does not
depend on $k$. Noting that $\xi_2 \to 0$ for $\Gamma \to 0$, 
the function becomes
\begin{equation}
\Sigma \simeq  A(\xi_1) \xi_2,  
\label{case1} 
\end{equation}
where $A(\xi_1)$ represents the direct contribution of $\Gamma$ that 
is not expressed through the dependence of $T$ on $\Gamma$.

Here, we consider $A(\xi_1)$ on the basis of the theory determining the 
behavior of shear stress
near the critical state for a dense colloidal suspension \cite{OtsukiSasa}. 
According to this theory, the critical behavior is described by an order 
parameter equation that is similar to the Ginzburg-Landau equation for
magnetization under a magnetic field. Assuming that this description
is valid for the present problem, we write 
\begin{equation}
c_0 (\nu-\Rc) A(\xi_1)- c_1 A(\xi_1)^3 =\xi_1
\label{GL}
\end{equation}
with numerical constants $c_0$ and $c_1$. Note that, in cases of
magnetic materials, $A(\xi_1)$ and $\xi_1$ correspond
to magnetization and a magnetic field, respectively.
Because the first 
term vanishes at the critical volume fraction, we obtain  
\begin{equation}
A(\xi_1)\simeq \xi_1^{1/3}.
\label{amp}
\end{equation} 

Following these assumptions, we can calculate the exponents $\alpha$.
Concretely, Eq. (\ref{case1}) with  Eq. (\ref{amp}) becomes 
\begin{equation}
\Sigma \simeq \xi_1^{1/3} \xi_2.
\label{amp2}
\end{equation}
Combining Eq. (\ref{amp2}) with Eq. (\ref{eb:dless}),  we derive
\begin{equation}
T \simeq a^2 m^{1/7} \eta^{6/7} \gamma^{8/7} .
\end{equation}
The substitution of this into Eq. (\ref{amp2}) with 
Eqs. (\ref{xi1}) and (\ref{xi2}) yields
\begin{equation}
\Sigma \simeq \left( \frac{\eta}{\sqrt{km}} \right)^{9/7}\Gamma^{5/7}.
\label{scaing:1:th}
\end{equation}
Thus, we obtain  $\alpha=5/7$ in Eq. (\ref{scaling:1}), 
which is consistent with the numerical experiment
as shown in Fig. \ref{fig:sg}.
In a similar manner, we obtain
\begin{equation}
\Pi \simeq \left( \frac{\eta}{\sqrt{km}} \right)^{10/7}\Gamma^{4/7}
\label{scaing:2:th}
\end{equation}
under the assumption that $\Pi(\xi_1,\xi_2) \simeq \xi_2$. 
This assumption implies  that $\Pi$ does not depend on $\xi_1$ 
because the normal stress is not directly influenced by the shear rate. 
As shown in Fig. \ref{fig:pg}, $\beta=4/7$ in Eq. (\ref{scaling:2}) 
is consistent with the numerical experiment.


\section{Interpretation of Eq. (\ref{hatano:power})}
We next study the power-law behavior observed in  the system 
under constant pressure on the
basis of the results obtained above. First, from Eqs. (\ref{scaling:1}) 
and (\ref{scaling:2}), which are valid at the critical volume fraction $\Rc$,
we derive Eq. (\ref{hatano:power}) with 
\begin{equation}
\phi=\frac{2(\alpha-\beta)}{2-\beta}.
\label{phi:rel}
\end{equation}
Using the values  $\alpha=5/7$ and $\beta=4/7$, we obtain $\phi=1/5$.
This value of $\phi$ is consistent with the numerical 
experiment. However, it should be noted that the scaling relation 
is obtained at the critical volume fraction, not for 
systems under constant pressure. 

In order to discuss quantitatively the behavior of the system under 
constant pressure, we denote the volume fraction and the shear stress 
measured on this system as $\nu=g_\nu(\Gamma,\Pi)$ and 
$\Sigma=g_\sigma(\Gamma,\Pi)$. 
We then wish to determine these functions from Figs. \ref{fig:sg} and 
\ref{fig:pg}. First, the point $(\Gamma,\Pi)$ in Fig. \ref{fig:pg} 
determines the volume fraction uniquely. We assume here that this 
volume fraction is realized in the system under constant pressure 
$\Pi$ with shear rate $\Gamma$ and that the shear stress in the system
is determined by using the volume fraction in Fig. \ref{fig:sg}.
Note that this assumption was  confirmed directly by a numerical experiment
for a constant pressure system whose size is close to that of the system 
with the Lees-Edwards boundary conditions. 
Based on  this assumption, we determine the volume fraction 
as a function of $\Gamma$ at a constant  pressure. In addition,
using this, we can obtain the dependence of the shear stress on 
$\Gamma$ at a constant  pressure. Hence, we obtain $\sigma/p$ as 
a function of $I$
under constant $\Pi$. For reference, we plot the lines 
$I={\rm const.}$ in Fig \ref{fig:pg}. As an example in Fig. \ref{fig:pg},
let us consider the case that $\Pi=10^{-3}$ in which the volume fraction 
is larger than the critical one for a sufficiently small $\Gamma$.
{ This case corresponds to the regime  $ I \le  10^{-3}$ and 
$\nu \ge 0.650$. Then, from Fig. \ref{fig:sg}, we find that the 
shear stress remains almost constant. Next,} 
in the interval $10^{-3} \le I  \le 10^{-1}$, the states with $\Pi=10^{-3}$ 
are close to the critical line. Thus, in this regime, it is expected that 
$\sigma/p$ behaves as that in the critical line, and the scaling behavior 
given in Eq. (\ref{hatano:power}) is observed approximately. 

Generalizing the above discussion, we expect the typical dependence  
of $\sigma/p$ on $I$ as  follows:
\begin{eqnarray}
\sigma/p &\approx&  {\rm const.} \quad {\rm for} \quad  I \ll I_0,
\label{sp:re0} \\
         &\approx&  I^{1/5}      \quad {\rm for} \quad  I_0 < I < I_1,  
\label{sp:re1} 
\end{eqnarray}
when spatially homogeneous shear flow is realized. Note that $I_0$ and 
$I_1$ are dependent on the pressure. For example, for states with  extreme pressures, such relations would not be observed. 
Since we wish to know the extent to which this approximate power-law relation
holds,  in Fig. \ref{fig:hatano}, 
we include the constant pressure data $(I, \sigma/p)$ obtained
from Figs. \ref{fig:sg} and  \ref{fig:pg}. 
As expected from the above consideration, 
the power-law behavior is observed for the case 
in which $\Pi=1.25\times 10^{-3}$. Furthermore, 
the system obeys the power-law regime
even for the case $\Pi=5.4 \times 10^{-5}$ in which the line is 
located below
the critical states in Fig. \ref{fig:pg}. 
We do not understand the reason why the power law
regime is so wide. 


\section{Concluding remark}

%
%

We have presented numerically and theoretically the scaling relations
given in Eqs. (\ref{scaling:1}) 
and (\ref{scaling:2}) for the system with the Lees-Edwards boundary 
conditions. From these new scaling relations, we also 
have an interpretation of the result observed in the system under 
constant pressure. The result is summarized in Eqs. (\ref{sp:re0}) 
and (\ref{sp:re1}). 

%
%

As far as we know, few experimental results exist in this regard. We 
expect 
that the power-law behaviors given in Eqs. (\ref{scaling:1}) and 
(\ref{scaling:2}) are observed in systems with constant volume, 
e.g., by operating a rotating Coutte-flow system \cite{fn:cv}. 
 With regard to Eq. (\ref{sp:re0}) that is valid in the very low
shear rate regime, we conjecture that a thermal activation process, 
which might lead to the logarithmic dependence of $\sigma/p$ on $I$,
occurs in this regime. Note that such behavior is ubiquitous in shear 
flow and sliding friction \cite{log}. 
Furthermore, the result reported in Ref. \cite{cross}  
might be related to 
Eq. (\ref{sp:re1}). We hope that more intensive experimental studies 
will be performed in this regard. 

%
%

\begin{figure}[htbp]
\begin{center}
\includegraphics[height=15em]{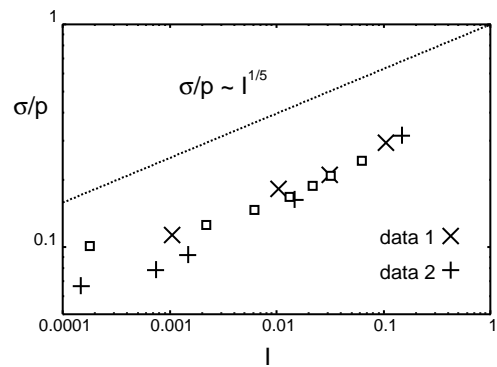}
\caption{
{ 
  $\sigma/p$ as a function of $I$ for the model described in 
the caption of Fig. \ref{fig:hatano} with a Herzian type interaction
 $k' (1-\delta r/(r_1+r_2))^{3/2}$. 
The parameter values are as follows: $L_x/a=L_y/a=25$, $N=10800$, and 
$\eta \sqrt{a}/\sqrt{k'm}=1.0$. $pa^2/k'=3.75\times 10^{-5}$ (data 1),
$1.92 \times 10^{-3}$ (data 2). Furthermore, the square symbols 
represent the constant pressure data  obtained 
from Figs. \ref{fig:sg:Hertz}  and  \ref{fig:pg:Hertz}, where 
the interaction force is $k'' ((r_1+r_2)-\delta r/)^{3/2}$ and
$p\sqrt{a}/k''=6.625\times 10^{-4}$.
}
}
\label{fig:hatano:Hertz}
\end{center}
\end{figure}

\begin{figure}[htbp]
\begin{center}
\includegraphics[height=15em]{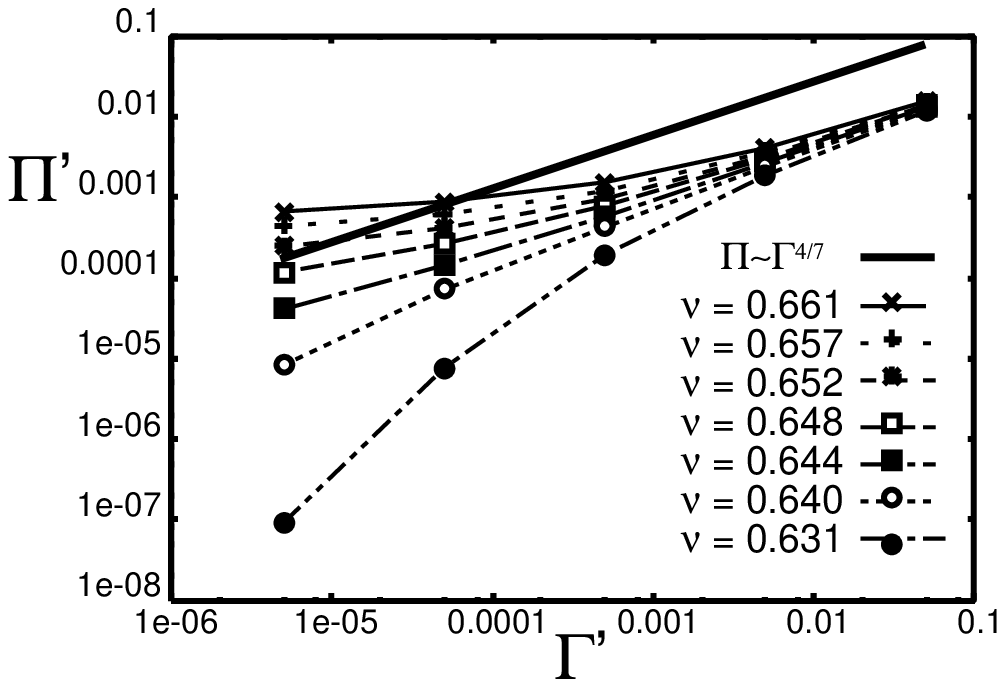}
\caption{
{ 
Dimensionless shear stress $\Pi'$ as a function of 
the dimensionless shear rate $\Gamma'$ for several values of $\nu$
with the interaction force $k'' ((r_1+r_2)-\delta r)^{3/2}$.
The dimensionless shear stress and the dimensionless shear rate 
are defined as $\Pi'=p \sqrt{a} / k''$
and $\Gamma'=\gamma \sqrt{m/(k'' \sqrt{a})}$, respectively.
The parameter values are as follows:  $N=1728$, and 
$\eta/\sqrt{k''m\sqrt{a}}=1.0$.
}
}
\label{fig:pg:Hertz}
\end{center}
\end{figure}

{Related to experimental studies, one may be interested in
the dependence of our result on the choice of the model 
we investigate.  For example, one may choose the Herzian 
type as an alternative for the interaction force. Indeed, 
such a model dependence has been discussed in the case
of zero-temperature and zero applied stress \cite{Nagel}. 
In our problem,  it is highly  expected that there is a  critical 
state at which rheological properties exhibit power-laws.
However, it is not evident that the exponents remain the same
values for the model with a Herzian type interaction.

\begin{figure}[htbp]
\begin{center}
\includegraphics[height=15em]{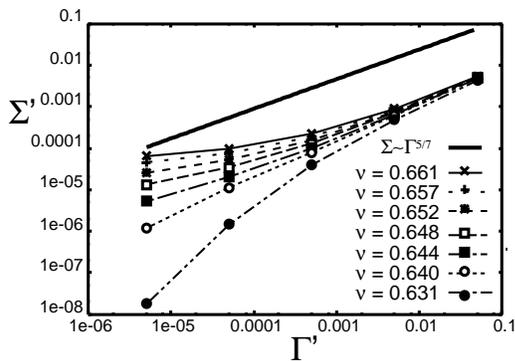}
\caption{
{ 
Dimensionless shear stress $\Sigma'$ as a function of 
the dimensionless shear rate $\Gamma'$ for several values of $\nu$
with the interaction force $k'' ((r_1+r_2)-\delta r)^{3/2}$.
The dimensionless shear stress and the dimensionless shear rate 
are defined as $\Sigma'=\sigma \sqrt{a} / k''$
and $\Gamma'=\gamma \sqrt{m/(k'' \sqrt{a})}$, respectively.
The parameter values are as follows:  $N=1728$, and 
$\eta/\sqrt{k''m\sqrt{a}}=1.0$.
}
}
\label{fig:sg:Hertz}
\end{center}
\end{figure}

In order to consider the model dependence explicitly, we 
demonstrate the result of numerical experiments in Figs. 
\ref{fig:hatano:Hertz}, \ref{fig:sg:Hertz}, and 
\ref{fig:pg:Hertz}. It is seen that the exponent of the 
system under constant pressure does not deviate so much 
from 1/5, while  the exponents $\alpha$ and $\beta$ seem 
to change slightly. We do not have a theoretical 
understanding  for these values yet, because the time scale 
related to the particle collision is not directly determined 
from model parameters. It might be important to develop a 
theory by using a more physical time scale such as the 
collision interval.  }

%
%

Finally, in our theoretical argument, Eq. (\ref{GL}) plays an essential role. 
As in the case of dense colloidal suspensions \cite{OtsukiSasa}, one may investigate 
the pair-distribution function of granular systems in order to derive the 
order parameter equation. The establishment of  
a complete theory in which the scaling relations are derived from a microscopic model is an important topic for future studies. 

The authors thank H. Hayakawa, N. Mitarai, and S. Tatsumi 
for their useful comments on this work. 
This work was supported by a grant from the Ministry of 
Education, Science, Sports and Culture of Japan (No. 16540337).


\begin{thebibliography}{10}

\bibitem{glass1}
L. Berthier and J. -L. Barrat, J. Chem. Phys. {\bf 116}, 6228 (2002). 

\bibitem{glass2}
J. Rottler and M. O. Robbins, Phys. Rev. E {\bf 68}, 011507 (2003).

\bibitem{glass3}
F. Varnik, L. Bocquet, and J. -L. Barrat, J. Chem. Phys. {\bf 120}, 2788
(2004).

\bibitem{Nagel}
C. S. O'Hern, L. E. Silbert, A. J. Liu, and  S. R. Nagel, 
Phys. Rev. E {\bf 68}, 011306 (2003). 

\bibitem{review}
GDR Midi, Europhys. J. E {\bf 14}, 341 (2004).

\bibitem{granular}
N. Xu and C. S. O'Hern, Phys. Rev. E {\bf 73},  061303 (2006). 

\bibitem{hatano}
T. Hatano, in preparation. 

\bibitem{fn:hatano}
Eq. (\ref{hatano:power}) is obtained when a tangential force between 
interacting particles is ignored. 
For general cases where the tangential force is taken into account, 
the left-hand side of Eq. (\ref{hatano:power}) is modified as $\sigma/p-\mu_0$,
where $\mu_0$ corresponds to the maximum static friction constant. The value 
of the exponent $\phi$ is identical to that in the case of Eq. 
(\ref{hatano:power}). The details will be reported elsewhere. 

\bibitem{Evans}
D. J. Evans and G. Morris, {\it Statistical mechanics of Nonequilibrium
Liquids}, (Academic, London, 1990). 



\bibitem{Campbell}
C. S. Campbell, J. Fluid Mech. {\bf 465}, 261  (2002).
 

\bibitem{fn:power2}
A power-law behavior similar to that in Eq. (\ref{scaling:1}) has been 
reported in the two-dimensional system \cite{granular}, though 
the critical state is not focused on.

\bibitem{Nature}
A. Liu and S. Nagel, Nature (London) {\bf 396}, 21  (1998). 

\bibitem{Bagnold}
R. A. Bagnold, Proc. R. Soc. London A {\bf 225}, 49  (1954).

\bibitem{Mitarai}
N. Mitarai and H. Nakanishi, Phys. Rev. Lett. {\bf 94}, 128001  (2005).

\bibitem{OtsukiSasa}
M. Otsuki and S. Sasa, e-print cond-mat/0511111.

\bibitem{fn:cv}
We  confirmed that a critical state was observed in the system
with walls consisting of the same kind of particles as those in the bulk, 
by which the volume is maintained constant. 

\bibitem{log}
C. Marone, Ann. Revs. Earth and Plan. Sci. {\bf 26}, 643 (1998);
F. Heslot, T. Baumberger, B. Perrin, B. Caroli, and C. Caroli, Phys. Rev. E {\bf 49}, 4973  (1994).

\bibitem{cross}
M. L. Blanpied, T. E. Tullis, and J. D. Weeks, Geophys. Res. Lett. 
{\bf 14}, 554 (1987).

\end{thebibliography}
\end{document}